\documentclass[conference]{IEEEtran}
\IEEEoverridecommandlockouts
\usepackage{cite}
\usepackage{amsmath,amssymb,amsfonts}
\usepackage{graphicx}
\usepackage{textcomp}
\usepackage{xcolor}
\usepackage{booktabs}
\usepackage{subcaption}
\usepackage{algorithm}
\usepackage{algpseudocode}

\def\BibTeX{{\rm B\kern-.05em{\sc i\kern-.025em b}\kern-.08em
    T\kern-.1667em\lower.7ex\hbox{E}\kern-.125emX}}
\begin{document}

\title{D2Q Synchronizer: Distributed SDN Synchronization for Time Sensitive Applications \\
}

\author{
    \IEEEauthorblockN{Ioannis Panitsas, Akrit Mudvari, Leandros Tassiulas}
    \IEEEauthorblockA{
        Department of Electrical and Computer Engineering, Yale University
    }
}

\maketitle

\begin{abstract}
In distributed Software-Defined Networking (SDN), distributed SDN controllers require synchronization to maintain a global network state. Despite the availability of synchronization policies for distributed SDN architectures, most policies do not consider joint optimization of network and user performance. In this work, we propose a reinforcement learning-based algorithm called \textit{D2Q Synchronizer}, to minimize long-term network costs by strategically offloading time-sensitive tasks to cost-effective edge servers while satisfying the latency requirements for all tasks. Evaluation results demonstrate the superiority of our synchronizer compared to heuristic and other learning policies in literature, by reducing network costs by at least 45\% and 10\%, respectively, while ensuring the QoS requirements for all user tasks across dynamic and multi-domain SDN networks.

\end{abstract}


\section{Introduction}

Software-Defined Networking (SDN) introduced a groundbreaking shift in network architecture by decoupling the control from the data plane, which were traditionally tightly integrated into the same networking device \cite{b1}. With this separation and the  centralization of the control plane, standardized interfaces were introduced to transport control actions and data, enabling programmable control over the underlying network infrastructure. Leveraging this architecture, network operators have deployed a range of SDN applications designed to enhance the performance and security of the network, including load balancing, dynamic routing, and software-based firewall applications. These  applications are hosted on commercial off-the-shelf commodity servers known as \textit{SDN controllers} and enable network programmability and reconfigurability, thereby offering network administrators increased flexibility to monitor, control, and secure their network. 

Although this centralized architecture offers substantial benefits, a distributed SDN architecture with multiple independent controllers managing individual network domains has been proposed to enhance scalability and fault tolerance \cite{b2}. These controllers operate in a \textit{logically-centralized} manner by periodically exchanging their network states to create a consistent view of the network, a process known as \textit{controller synchronization}. 
\begin{figure}[ht]
  \centering
  \includegraphics[width=0.45\textwidth]{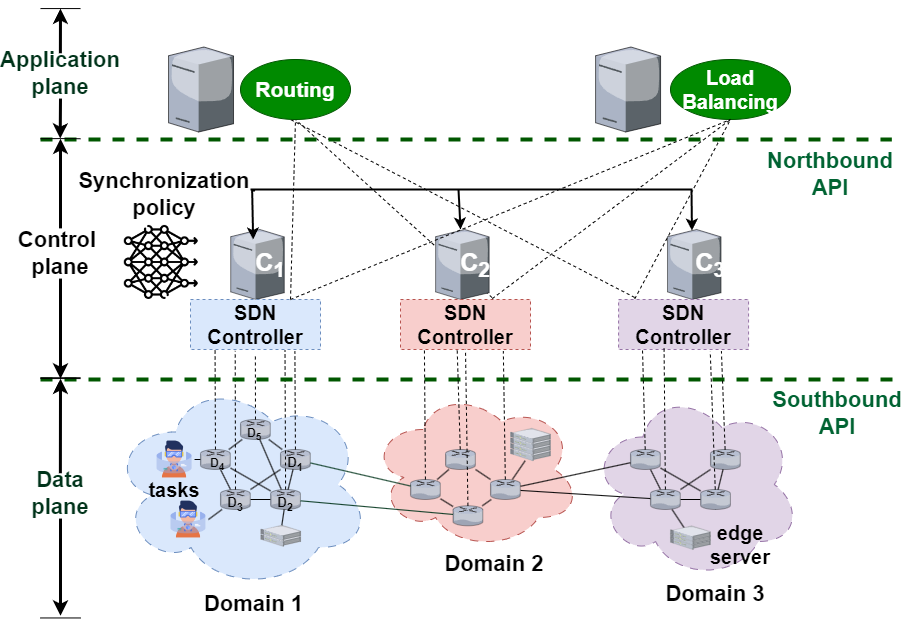}
  \caption{A learning-based synchronization policy in $C_1$ determines which controllers to exchange state within each period to maintain a global network view.}
  \label{fig:sdn_architecture}
\end{figure}
In large-scale networks, complete synchronization among controllers is often cost-prohibitive due to the significant communication overhead in the control plane. As a result, partial synchronization is employed, accepting temporary inconsistencies in the controllers’ states whenever such inconsistencies are deemed acceptable, a concept commonly referred to as the \textit{eventual consistency model} \cite{b3}.

Recent research works have focused on developing learning-based algorithms to maintain consistent network state \cite{b4}, \cite{b5}, \cite{b6}, \cite{b7}, \cite{b8}. The authors in \cite{b4}, proposed an algorithm for determining the optimal frequency of exchanging synchronization messages among controllers for shortest path routing and load balancing in SDN applications. The authors in \cite{b5} proposed a Deep Reinforcement Learning (DRL) algorithm to minimize communication latency and waiting time for inter-domain routing tasks by selectively synchronizing the optimal subset of SDN controllers. In \cite{b6}, the authors introduced a DRL-based synchronization policy for inter-domain routing tasks. In \cite{b7}, DRL and transfer learning synchronization policies were developed for inter-domain routing and load balancing, respectively. Finally, in \cite{b8}, a joint controller synchronization and placement RL algorithm was proposed and evaluated in inter-domain routing and load balancing SDN applications. 

Despite these research efforts, we have identified several limitations in the existing works. Firstly, most approaches are limited to optimizing a single performance metric, failing to address more complex and realistic objectives \cite{b4}, \cite{b5}, \cite{b6}, \cite{b7}. Secondly, they predominantly emphasize either user-centric \cite{b5} or network-centric metrics \cite{b6}, \cite{b7} without considering both simultaneously. Thirdly, they primarily concentrate on a limited set of SDN applications \cite{b4}, \cite{b5}, \cite{b6}, \cite{b7}, \cite{b8}, such as shortest path routing; however, there is a critical need to address a broader range of use cases, especially in Next-Generation (NextG) networks, where numerous new applications are expected to emerge with stringent QoS requirements, such as augmented and virtual reality applications. Finally, there is a notable lack of comparison with state-of-the-art  synchronization policies, as well as comprehensive evaluations in dynamic and multi-domain SDN networks \cite{b5}, \cite{b6}, \cite{b7}, \cite{b8}. 

To address these limitations, we propose a novel synchronization algorithm that can be deployed in any distributed SDN controller to facilitate intelligent synchronization decisions, enabling long-term network cost optimization while meeting the strict latency requirements of all tasks and maintaining satisfactory user performance. We compare our \textit{D2Q Synchronizer} with heuristic and learning-based state-of-the-art synchronizers in various dynamic and multi-domain SDN networks. In all scenarios, it outperformed heuristic policies by at least 45\% and intelligent synchronizers by at least 10\% in minimizing network costs. Finally, in addition to cost minimization, our synchronizer increased both the total number of latency-compliant paths and the number of correct server allocations for task offloading compared to the other synchronizers.

\section{Motivation and System Description}

In this study, our primary objective is to develop an intelligent synchronization policy with a limited synchronization rate, aiming to minimize the number of control plane state exchanges while maintaining satisfactory user and network performance. The policy will assist the controller in approximating the long-term global network state as closely as possible to the ground-truth state by carefully selecting which controllers to synchronize at each decision step. The development of this algorithm is driven by two primary reasons. 

First, we aim to minimize the number of control plane messages exchanged between the controllers, because they carry large amounts of data \cite{b1}, \cite{b2}. Although the user plane generates most of the network traffic \cite{b1}, reducing unnecessary control plane messages is crucial, particularly in multi-domain NextG SDN networks. For instance, in a network with $N$ SDN controllers that are connected with a mesh topology, the number of synchronization messages grows at a rate of $\mathcal{O}(N^2)$, making fully consistent algorithms difficult to scale in large networks. In addition to this, each network state update requires transmitting substantial data within a short time period, which necessitates a large pool of communication and computation resources \cite{b1}, \cite{b2}. 

The second motivation arises from the extensive deployment of wireless SDN controllers in NextG networks, spanning from the radio access network to the core network \cite{b15}, \cite{b16}. Many of these sections include wireless segments, meaning SDN controllers must manage domains that rely on wireless channels. These wireless channels, which either connect the controllers or the control with the data plane, are vulnerable to disruptions caused by factors such as poor signal quality, the mobility of SDN controllers, or adversarial actions like jamming attacks. Therefore, implementing a policy that limits synchronization among controllers can help sustain robust user and network performance, even in uncertain and dynamic network environments.

\subsection{Distributed SDN Environment}

Our study focuses on a distributed SDN environment, where SDN controllers ($C$) and data plane devices ($D$) are interconnected through wireless links. Each network domain includes SDN-enabled data plane devices, such as programmable switches and routers, as well as edge servers for task offloading and processing. These domains are interconnected through gateway nodes with inter-domain links. Distributed SDN controllers have an up-to-date view of their own domains by continuously monitoring intra- and inter-domain link and node-level statistics, including link latency, throughput, and computing resource utilization. The network domains are highly heterogeneous and dynamic, with varying numbers of routing and computation devices to meet user traffic demands. The available capacity of intra- and inter-domain links is constantly evolving due to the background traffic in each domain, as well as the available computing resources in edge servers. In this study, we assume that each edge server $e \in E$, from a set of servers $E$, is assigned a dynamic cost variable denoted as $c_e$, which represents the processing cost of completing a task on that server. Finally, we assume that the queue waiting time is uniform across all routing devices.

\subsection{Application of Interest}

In this work, we focus on strategically offloading tasks to the most cost-efficient servers while ensuring that these tasks meet their latency deadlines. More specifically, within the domain (controller) where the synchronization policy is deployed, user-generated tasks are collected and forwarded from the data plane devices to the domain's SDN controller, which selects the optimal paths and the appropriate edge servers for task offloading. In this setup, the objective of the SDN controller is twofold: first, to calculate end-to-end paths that may span multiple network domains and satisfy the latency requirements of user tasks; and second, to select servers that minimize total network costs. While this task may seem straightforward, it presents significant challenges due to limited synchronization. The SDN controller lacks a global view of the entire network and can only receive network state information from a subset of neighboring SDN controllers. This means that, the correct calculation of paths and server selections is highly dependent on the synchronization rate, due to the dynamic nature of each network domain’s topology, as well as the varying link capacities and server costs. For instance, higher synchronization rates will assist the controller to calculate more accurate paths and select optimal servers, but at the expense of increased control plane communication. Conversely, lower synchronization rates result in less accurate calculations. Therefore, a policy that autonomously and intelligently selects the optimal subset of SDN controllers is needed to balance communication costs with network and user performance.

\subsection{Problem Description}

Before describing the problem, we define the unit that quantifies the synchronizable domain information and the synchronization budget.

\noindent \textbf{Definition 1.} \textit{ A synchronization control message (SCM) encapsulates the intra-domain topology, intra- and gateway link delays, and edge server costs. It reflects the synchronizable domain information that needs to be exchanged between domain controllers.}

\noindent \textbf{Definition 2.} \textit{ The maximum number of SCMs that can be exchanged between SDN controllers in a time period $\tau$ is defined as the synchronization budget ($SB$).}

Another key aspect to address is the method by which the SDN controller determines the optimal network path and selects an appropriate server for each task.

\noindent \textit{Server Selection}: The server selection for the SDN controller is straightforward, it selects the server with the minimum cost (based on its global network state) across the entire network:
\begin{equation}
e^* = \mathop{\arg\min}_{e \in E} c_e
\end{equation}

\noindent \textit{Path Calculation}: Paths are computed using Dijkstra's algorithm, where the weights of the network correspond to the transmission latency across both intra-domain and inter-domain links. Therefore, to determine the optimal path (in terms of latency) \( p^*_{k,e} \in P \) from a set of candidate paths \( P \), starting from an initial node \( k \) and ending at an edge server \( e \in E \), the SDN controller uses the following formula:
\begin{equation}
p^*_{k,e} = \mathop{\arg\min}_{p \in P} d(p) 
\end{equation}

\noindent where \( d(p) \) denotes the total transmission latency due to the intra and inter-domain link latencies from the selected path $p$. 

Consider a distributed network architecture as illustrated in Figure 1, with \( N \)  SDN controllers. Let's assume that the synchronization policy is deployed in \( C_1 \) controller, and the controller can communicate with only $SB$ controllers during each time period $\tau$. During this time period, tasks, denoted as \( J \), are generated by the users or sampled from the queues of the previous time step $\tau -1$. Each task $j \in J$ has a predefined latency requirement denoted as $l_j$. The \( C_1 \) controller, based on its current network state, selects the optimal set of edge servers \(\hat{E_*} \subseteq E \), that satisfy the latency constraints for all tasks $J$, as shown in Eq. (3) and (4).

\begin{equation}
     \hat{E_*} = \min_E \sum_{j=1}^{J} c_{e_j}
\end{equation}
\begin{align}
    \text{subject to:} & \quad d(p) \leq l_j, \quad \forall j \in J   
\end{align}

\noindent Due to limited synchronization, however, the controller may not always select the globally optimal set of edge servers \( E_* \subseteq E \). Thus, the policy must strategically guide the \( C_1 \) controller to synchronize with only a subset of SDN controllers (constrained by \( SB \)) that will yield long-term cost minimization benefits. In a sense, the policy should autonomously understand which domains are more dynamic at every time step (in terms of topology, link latencies, and server cost variations, as all of these variables are dynamic) and, based on these insights, to synchronize the appropriate SDN controllers. To approximate the optimal synchronization policy, we leverage the power of RL algorithms, which are well-suited for solving such complex sequential decision-making problems.

\subsection{MDP Formulation}
\noindent We formulate our controller synchronization problem as a MDP \cite{b10} where:

\begin{itemize}

    \item \( S \) is our finite state space, represented by a one-dimensional vector \( \mathbf{s} \in \mathbb{R}^N \), and denotes the number of time periods elapsed since the last  synchronization of each SDN controller.  This representation enables us to monitor the frequency of updates from the controllers, providing insights into the synchronization policy.

    \item \( A \) is our finite action space, represented by a one-dimensional vector \( a \in \{0, 1\}^N \), where each element \( a_i \) corresponds to the decision made for the i-th controller. Specifically, \( a_i = 1 \) indicates that the i-th controller's SCM is to be exchanged during the current time step, while \( a_i = 0 \) shows that the i-th controller's SCM will not be exchanged.

    \item Our reward function \( R(s,a) \) reinforce the selection of SDN controllers that will help to minimize the long-term network costs as described in Equation (3). If the selected controllers are suboptimal in terms of cost minimization or violate the latency requirements of tasks, the reward function should return a penalty that guides the agent to avoid these actions. For instance, a sub-optimal selection of controllers for synchronization that increases the number of tasks with latency violations should be penalized with a negative reward. Conversely, if the selected controllers reduce task latency violations but result in more sub-optimal task allocations to edge servers, the agent should receive a smaller negative reward compared to the first case. Therefore, our reward function is defined as follows:
    \begin{equation}
         R(s,a) = \sum_{j=1}^{J} U(j,p(\hat{e}),e^*)
    \end{equation}
    
\begin{equation}
    \begin{split}
        U(j,p(\hat{e}),e^*) = 
        \begin{cases} 
        0, & \text{if } d(p(\hat{e})) < l_j \\
           & \text{and } \hat{e} = e^* \\[5pt]
        -\lambda \left| c_{\hat{e}} - c_{e^*} \right|, & \text{if } d(p(\hat{e})) < l_j \\
           & \text{and } \hat{e} \neq e^* \\[5pt]
        -r_1, & \text{otherwise}
        \end{cases}
    \end{split}
\end{equation}

    where $\lambda$ and $r_1$ are positive scalars, while $e^*$ and $\hat{e}$ represent the optimal cost-efficient server and the selected server for each task $j$, respectively. The values of $\lambda$ and $r_1$ were set to 80 and 10000 respectively.

\end{itemize}

\section{D2Q Synchronizer}

In this section, we provide the details of our \textit{D2Q Synchronizer}, our RL agent, which was utilized to solve the MDP defined previously. In general, in RL, an agent sequentially interacts with an environment, taking actions and receiving feedback in the form of rewards or penalties. The agent's objective is to learn a policy that maximizes the cumulative reward over time. In our case, the goal of the \textit{D2Q Synchronizer} is to find a sequence of actions within a finite discrete horizon $T$ such that the long-term accumulated reward is maximized. This will enable the SDN controller to correctly offload tasks to latency-compliant paths, minimizing the network operator's costs. While tabular - RL approaches can be very successful for small state-action spaces, in our problem the state-action space is extremely large. The state space grows as \(T^N\), while the action space grows as \(\binom{N}{SB}\). To overcome this issue, we employ DRL techniques that can handle large state-action spaces, such as playing Atari games \cite{b11}, and more specifically we use Double Deep-Q Networks (DDQN) \cite{b12} as the core for our agent.

\textit{D2Q Synchronizer Architecture}: The input to our \textit{D2Q Synchronizer} consists of the synchronization state of all controllers, while the output is the selected subset of controllers for synchronization. To accurately map and select the optimal action \( a \) in each state \( s \), we employ two Deep Neural Networks (DNNs): a main network for generating action selections and a target network for providing stable target values during training. 

\textit{Training the D2Q Synchronizer}: To train our synchronizer, we first created a range of simulated network environments with varying numbers of domains, data plane devices, service requests, and synchronization budgets, as detailed in Section IV.B. For each environment, we trained our agent and monitored its performance in minimizing accumulated network costs while ensuring that latency requirements for user requests were met. Both the training and exploitation phases of the agent occurred within the same evolving network over time.  For balancing the training-exploitation phase we utilized an $\epsilon$-greedy policy.  Further details of the training process are outlined in Algorithm 1.

\begin{algorithm}
\caption{Training D2Q Synchronizer}
\textbf{Input:} randomized parameters \(\theta\), \(\theta'\) for initializing the main and target DNN, learning rate \(\alpha\), batch size \(B\), epsilon decay rate \(\epsilon\), soft update rate for target network \(\kappa\), time horizon $T$ \\
\textbf{Output:} learned parameters \(\theta\) for DNN
\begin{algorithmic}[1]
\State Initialize: Main DNN with random parameters \(\theta\) and target DNN with parameters \(\theta' = \theta\)
\State Initialize: Experience Replay Buffer \(I\) for \((s, a, r, \phi)\) tuples to be stored for episodes \(E = 1, ..., M\)
\For{$E = 1, ..., M$} 
\State Set: Exploration probability \(p_e = \frac{1}{1+E/\epsilon}\)
\State Set: Initial state \(s_0\) to zero vector

    \While{$t \leq T$}
    \State Select action \(a_t\) using the epsilon-greedy policy: \Statex \hspace{2.7em} with probability \(p_e\), select a random action; \Statex \hspace{2.7em} otherwise, select \(a(t) = \arg\max_a Q(s(t), a; \theta)\)

    \State Obtain the rewards \(r_t\) from the \Statex \hspace{2.7em} simulated network environment using equation (6)
    \State Transition to the next observation \(\phi(t) = s(t+1)\)

    \State Store \((s(t), a(t), r(t), \phi(t))\) in \(I\)
    \State Sample a random mini-batch of size \(B\) from \(I\), \Statex \hspace{2.7em} each entry denoted by \((s_j, a_j, r_j, \phi_j)\), \(j = 1, ..., B\)
    \For{$j = 1, ..., B$}
        \State \(a'_j = \arg\max_A Q(s_j, A; \theta)\) 
        \State \(y_j = r_j + \gamma Q(\phi_j, a'_j; \theta')\) 
        \State Perform gradient descent step on \Statex \hspace{4em} \((y_j - Q(s_j, a_j; \theta))^2\) with respect to \(\theta\), to obtain \Statex \hspace{4em} \(\nabla_{\theta}L(\theta)\)
    \EndFor
    \State \(\theta \leftarrow \theta - \alpha \nabla_{\theta} L(\theta)\)
    \State \(\theta' \leftarrow \kappa \theta + (1 - \kappa) \theta'\) 
    \EndWhile
\EndFor
\end{algorithmic}
\end{algorithm}

\begin{figure*}[ht]
    \centering
    \begin{subfigure}[b]{0.29\textwidth} 
        \centering
        \includegraphics[width=\textwidth]{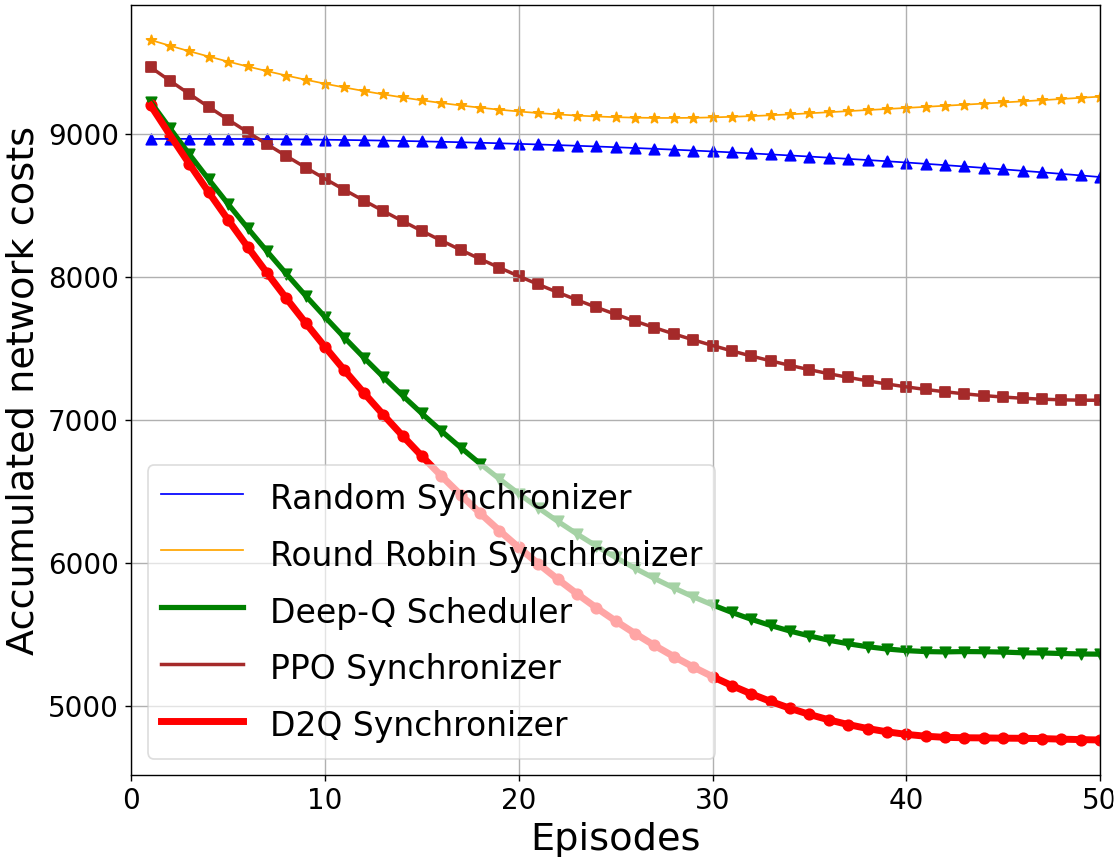}
        \caption{Reduction in network costs.}
        \label{fig:costs}
    \end{subfigure}
    \hspace{0.01\textwidth} 
    \begin{subfigure}[b]{0.29\textwidth}
        \centering
        \includegraphics[width=\textwidth]{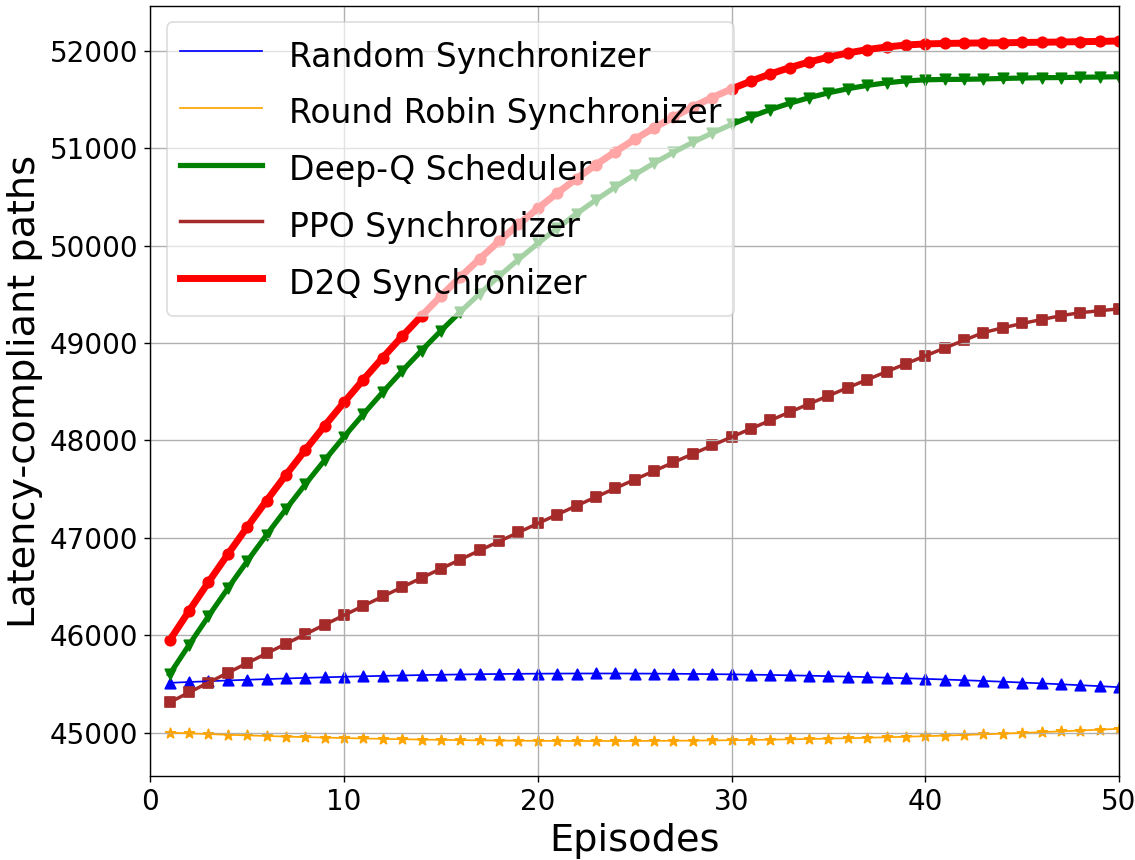}
        \caption{Increase in latency-compliant paths.}
        \label{fig:paths}
    \end{subfigure}
    \hspace{0.01\textwidth} 
    \begin{subfigure}[b]{0.30\textwidth}
        \centering
        \includegraphics[width=\textwidth]{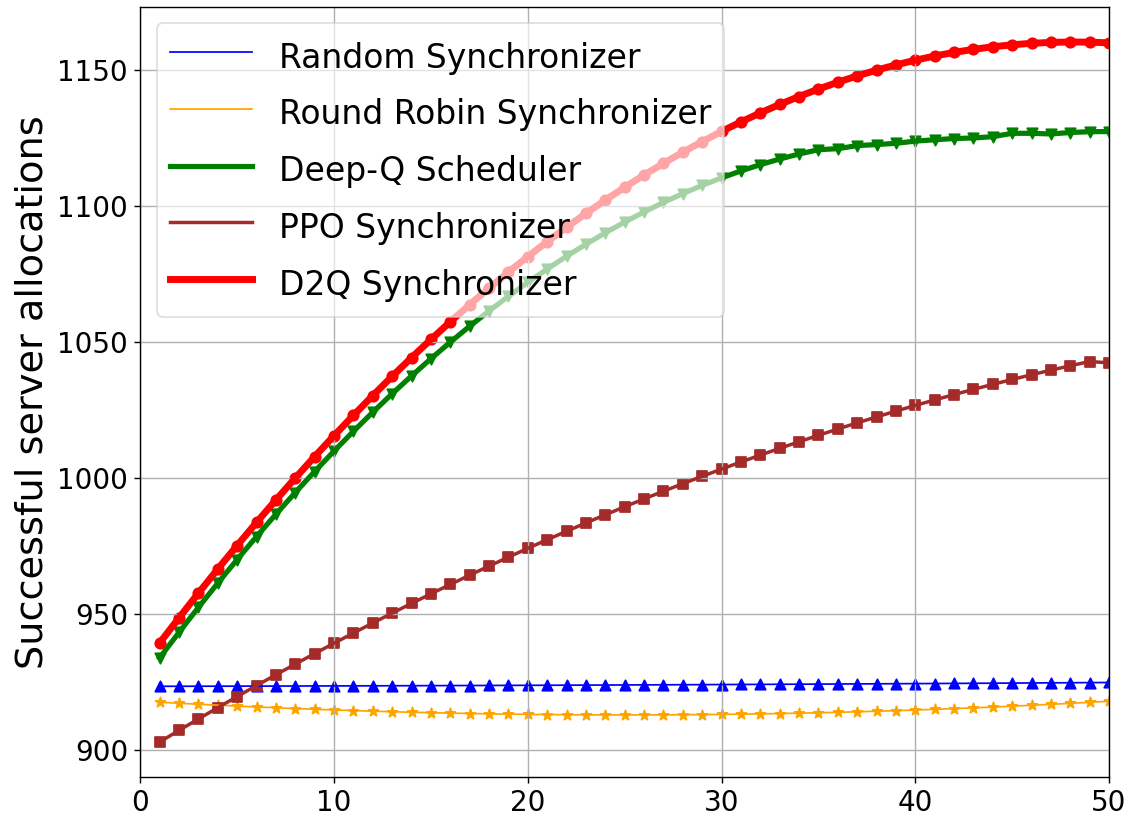}
        \caption{Increase in correct server allocations.}
        \label{fig:allocations}
    \end{subfigure}
    
    \caption{Comparison of accumulated costs, latency-compliant paths, and server allocations across all  synchronization algorithms.}
    \label{fig:combined}
\end{figure*}

\section{Evaluation}

\subsection{Synchronizers}
To evaluate the performance of our \textit{D2Q Synchronizer}, we implemented two well-known policies: the Random policy and the Round Robin policy, as implemented by the authors in \cite{b7} and \cite{b8}. Additionally, we implemented an intelligent RL-based synchronizer leveraging the Proximal Policy Optimization (PPO) algorithm \cite{b13}, referred to as the PPO Synchronizer. Finally, we compare our synchronizer with the state-of-the-art DQ-Scheduler \cite{b6}, that was fine-tuned for our task. Note that these algorithms do not create a fully consistent network state, enabling a fair comparison with our \textit{D2Q Synchronizer}.

\noindent \textbf{Random Synchronizer:} A controller synchronization policy that randomly synchronizes a subset of controllers $|SB|$ during each time period $\tau$. 

\noindent \textbf{Round Robin Synchronizer:} A controller synchronization policy that sequentially synchronizes subsets of SDN controllers at each time period $\tau$. 

\noindent \textbf{PPO Synchronizer:} An RL-based synchronization policy that leverages the PPO algorithm to intelligently synchronize SDN controllers. This method intelligently selects the SDN controllers with the highest priority for synchronization. In our implementation, the discount factor $\gamma$ was set 0.01, encouraging a more greedy synchronization strategy that prioritizes immediate over long-term rewards.

\noindent \textbf{DQ Scheduler:} An RL-based scheduler originally designed for interdomain routing tasks \cite{b6}, which was fine-tuned for our synchronization application by modifying its reward function to align with our objectives. The core implementation of this policy is based on Deep-Q Networks.
More details regarding the architecture of the RL-based synchronizers and the hyperparameters used are provided in Section IV.C.

\subsection{Network Settings}
\begin{enumerate}
    \item \textit{Network Topology of the Simulated SDN Network}: 
    For our evaluations, network topologies within each domain, as well as the inter-domain connections, were generated using the Erdős–Rényi model \cite{b9}. Intra-domain links were subject to potential failures, with a probability of \( p = \frac{1}{30} \) during each time period $\tau$. The algorithms were evaluated across various SDN networks, one network each time for training and inference, with the number of domains (\(N\)) ranging from 5 to 12, and each domain containing between 2 and 15 data plane devices $D$. Edge computational resources were distributed in each domain, with 4 servers and each assigned a randomly assigned cost (\(c\)) ranging from 20 to 100 each time period $\tau$. Finally, the synchronization budget (\(SB\)) was set between 2 and 8, for each network.
 
    \item \textit{Task Generation}: In each data plane device, user tasks are generated according to a Poisson distribution with a rate parameter (\(\lambda\)) ranging from 2 to 5, each with varying latency requirements. We evaluated two scenarios: tasks with stringent low latency requirements, categorized as low latency (10 ms), and tasks with more relaxed time constraints, categorized as mid latency (100 ms) \cite{b14}.
  
\end{enumerate}

\subsection{Synchronizers Settings}

Our \textit{D2Q Synchronizer}, along with the other policies, was fine-tuned through extensive hyperparameter optimization (Grid Search). A fair comparison across all the synchronizers was ensured by maintaining a uniform architecture consisting of two hidden layers with 64 neurons each, utilizing ReLU as activation function.  To enhance stability and convergence during training, we employ dropout regularization and an experience replay buffer \(I\) of size 40,000 to store the agent's experiences. For all the policies, we used a batch size (\( B \)) of 256, a learning rate ($\alpha$) of 0.01, a time horizon ($T$) of 500, an epsilon decay rate ($\epsilon$) of 25 and Adam optimizer.  Finally, the discount factor (\( \gamma \)) was set to 0.9 for all models, except for the PPO synchronizer, which was set to 0.01.

\subsection{Evaluation Results}

\begin{figure*}[ht]
  \centering
  \includegraphics[width=0.85\textwidth]{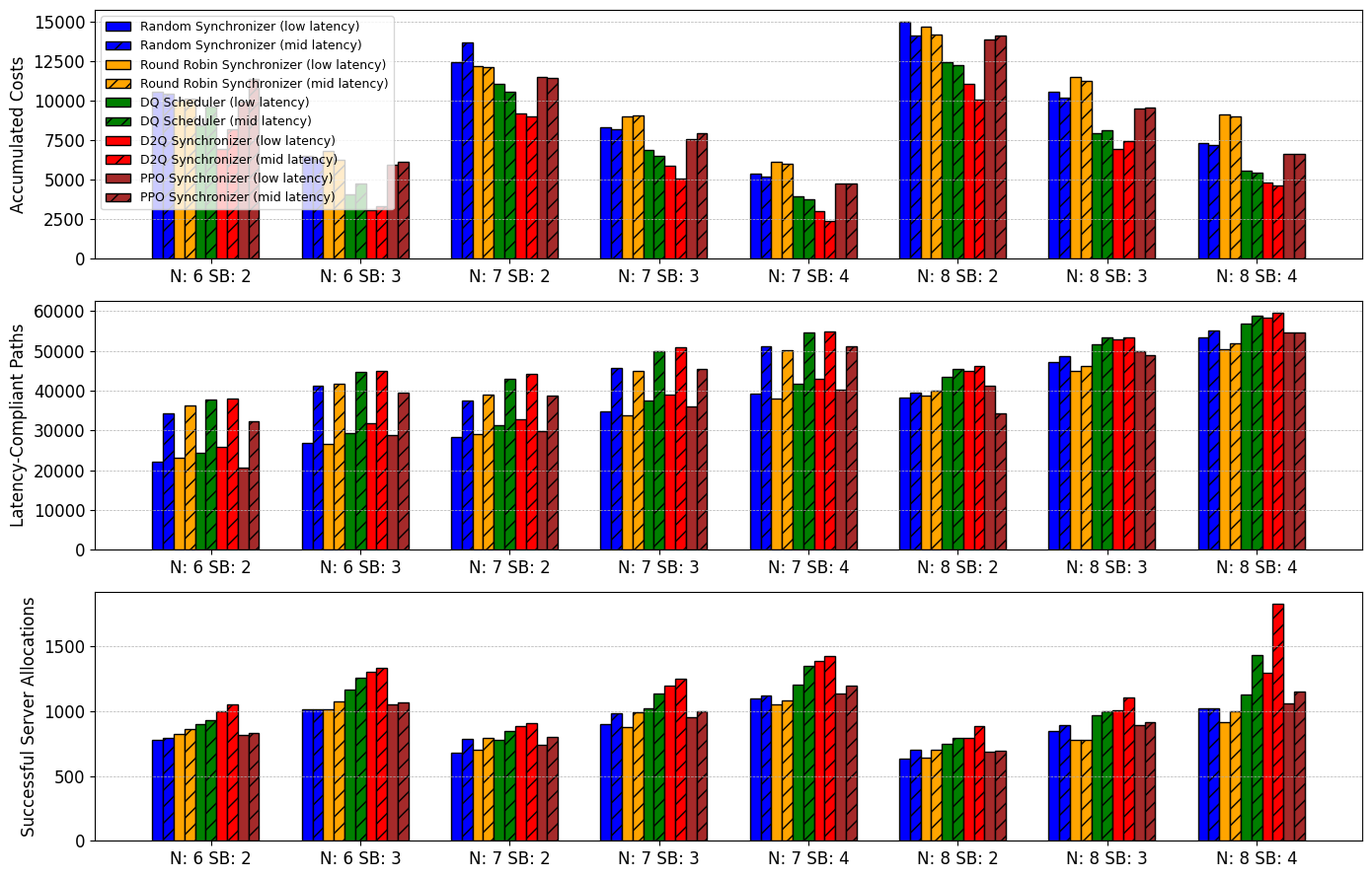} 
  \caption{Comparison of accumulated costs, latency-compliant paths, and server allocations across all synchronization algorithms under various SDN networks and latency requirements.}

  \label{fig:output}
\end{figure*}

The evaluation results of the \textit{D2Q Synchronizer} for one scenario with \( N = 7 \) domains and \( SB = 3 \) are presented in Fig.2, where Figs. 2a-2c show the performance in terms of the objective stated in Section II-C. Fig. 3 shows the performance of the \textit{D2Q Synchronizer} across different networks, with a varying number of domains, synchronization budgets, and tasks with different latency requirements.

\noindent \textit{1) Superiority of D2Q Synchronizer in minimizing costs:}

The evaluation results in Fig.2a confirm the superiority of the D2Q synchronizer in achieving long-term minimization of network costs. In particular, during the evaluation period of the last 25 episodes, where each episode lasted for \(T=1000\) time periods, the \textit{D2Q Synchronizer} outperformed the Random Synchronizer by 44.52\%; the Round Robin Synchronizer by 47.34\%; the PPO Synchronizer by 32.76\%; and the DQ Scheduler by 10.65\% in minimizing network costs.

\noindent \textit{2) Superiority of D2Q Synchronizer in secondary objectives:}

Despite our first objective being to minimize network costs by offloading user tasks to the most cost-efficient servers, the way we defined our reward function in Eq. (6) and (7) helped in maximizing the total number of latency-compliant paths as well as the number of optimal allocations to the most cost-efficient servers. In particular, the \textit{D2Q Synchronizer} outperformed the Random Synchronizer by 14.07\%; the Round Robin Synchronizer by 15.5\%; the PPO Synchronizer by 6.23\%; and the DQ Scheduler by 0.71\% in maximizing compliant paths, as illustrated in Fig. 2b. In addition, the \textit{D2Q Synchronizer} surpassed the Random Synchronizer by 24.57\%; the Round Robin Synchronizer by 25.79\%; the PPO Synchronizer by 12.18\%; and the DQ Scheduler by 2.68\% in maximizing optimal server allocations for offloading tasks, as illustrated in Fig. 2c.

\noindent \textit{3) Superiority of D2Q Synchronizer in various SDN networks and task deadlines:}
We evaluated the \textit{D2Q Synchronizer} across different SDN networks and latency requirements of tasks as described in Section IV.B, and in all these different settings, it outperformed all the baselines as illustrated in Fig. 3. Some interesting insights are that, first, higher values of $SB$ for a fixed value of $N$ will result in further improvements of the accumulated costs and the secondary objectives due to higher control plane communication. Secondly, less strict latency requirements of the tasks will have a greater impact on cost minimization as more latency-compliant paths and server allocations can be calculated correctly. Therefore, based on these observations, a network operator can fine-tune the $SB$ to balance control plane overhead with cost minimization.

\section{Conclusion}

In this paper, we studied the controller synchronization problem in distributed SDN for finding the optimal synchronization policy to minimize long-term network costs while jointly satisfying the QoS of user tasks. We proposed a new RL-based policy called the \textit{D2Q synchronizer} to solve the formulated MDP. Evaluation results demonstrated that our policy offers improved user and network performance compared with heuristics and state-of-the-art synchronizers.

\section*{Acknowledgment}

This research was in part supported by the Army Research Office W911NF-23-1-0088,  the National Science Foundation CNS-2146838 and CNS-2128530.

\vspace{12pt}
\color{red}

\end{document}